\documentclass[a4paper,fleqn,usenatbib]{mnras}

\usepackage[T1]{fontenc}
\usepackage{ae,aecompl}

\usepackage{graphicx}	
\usepackage{amsmath}	
\usepackage{amssymb}	
\usepackage{subcaption}
\usepackage{multirow}

\usepackage{newtxtext,newtxmath}

\title[Detectability of binary star companions from ETVs]{The Detectability of Binary Star Planetary and Brown Dwarf Companions From Eclipse Timing Variations}

\author[Getley et al.]{
A. K. Getley,$^{1}$\thanks{E-mail: \href{mailto:alan.getley@usq.edu.au}{alan.getley@usq.edu.au}}
B. Carter,$^{1}$
R. King$^{1}$
and S. O'Toole$^{2}$
\\
$^{1}$University of Southern Queensland, Centre for Astrophysics, Toowoomba Qld 4350 Australia\\
$^{2}$Australian Astronomical Optics, Macquarie University, PO Box 915, North Ryde NSW 1670 Australia
}

\date{Accepted XXX. Received YYY; in original form ZZZ}

\pubyear{2020}

\begin{document}
\label{firstpage}
\pagerange{\pageref{firstpage}--\pageref{lastpage}}
\maketitle

\begin{abstract}
In this paper, we determine the detectability of eclipsing binary star companions from eclipse timing variations using the Kepler mission dataset. Extensive and precise stellar time-series photometry from space-based missions enable searches for binary star companions. However, due to the large datasets and computational resources involved, these searches would benefit from guidance from detection simulations. Our simulations start with and benefit from the use of empirical Kepler mission data, into which we inject third bodies to predict the resulting timing of binary star eclipses. We find that the orbital eccentricity of the third body and the orbital period of the host binary star are the key factors in detecting companions. Target brightness is also likely to be a factor in detecting companions. Detectable third body masses and periods can be efficiently bound using just two equations. Our results enable the setting of realistic expectations when planning searches for eclipsing binary star planetary and brown dwarf companions. Our results also suggest the brown dwarf desert is real rather than observational selection.

\end{abstract}

\begin{keywords}
binaries: eclipsing
\end{keywords}



\section{Introduction}
For binary stars, eclipse timing variations (ETVs) measured from time-series photometry enable searches for the gravitational effect of additional, planetary or brown dwarf companions. Today the precise time-series observations needed come from space telescope missions such as Kepler \citep{prsa2011, slawson2011} and TESS \citep{ricker2014}. However, due to the large datasets and computational resources involved, searches for binary star companions need detection simulations to improve their efficiency. Realistic detection expectations save resources by directing searches to those systems where companions and their characteristics are most likely to be reliably obtained from the available dataset. A determination of the detectability of a companion reduces the rate of false positives and provides a check on the robustness of existing detections.

Over 2,000 eclipsing binary stars are listed in the Kepler Eclipsing Binary Star Catalog, which is the focus of this study as it alone provides such a large number of systems for ETV studies. Nevertheless, given that all observations will contain some level of random and systematic light curve errors, an understanding of the inherent capacity of the dataset to produce detections can assist with the identification of previously missed companions, and provide a check on known candidates. Thus it is important to understand the limitations on detections based on ETVs, and this can be done using simulations that introduce companions into a dataset of eclipsing binary star systems.

Planets with a large mass of $\sim$10 Jupiter Masses in long ($\sim$10 - 20 year) orbits can be detected with timing accuracies of $\sim$10s \citep{ribas2006}. Giant circumbinary planets can be detected through eclipse timing studies with timing precisions of between 0.1 and 1 second \citep{sybilski2010}. Numerical simulations performed by \cite{sybilski2010} indicate this required precision can be reached with the Kepler and the CoRoT missions. It is unlikely this precision can be achieved in practice. However, very little has been done to determine the practical limits of what has been observed and what third body masses may be too small to detect with these "real-world" observations. By using a binary star system that has been observed by Kepler and modelling the system in {\sc jktebop} \citep{southworth2004} we are able to estimate masses for the binary star components as well as orbital characteristics such as orbital period and inclination. By creating a model system based on these estimates we are then able to inject a third body with varying characteristics and run an eclipse time study on these simulated systems. 

In past papers we have presented evidence for a planetary mass third body orbiting KIC 5095269 found via an ETV study \citep{getley2017} as well as the stability of third bodies found around Kepler systems via ETV studies \citep{getley2020}. These papers naturally lead us to the question what the limits of ETV studies are when using "real-world" data (or Kepler derived jitter) as a base.

In this paper, we report the results of an eclipse time study on simulated systems using three different systems observed by Kepler as a base that have then been injected with third bodies. Therefore, we are able to report on the limits of detection using eclipse time variations using actual limits of the Kepler observations and variability inherent to the system. We are also able to report on what characteristics of third bodies may be detected or not within these limits. We performed this investigation using a mostly automated technique that is more widely applicable, as a manual process for thousands of systems is impractical unless there is a specific reason to look at a system manually (for example if another investigation into a specific system indicated a third body).

\section{Method}

\begin{figure}
    \centering
    \includegraphics[width=\linewidth]{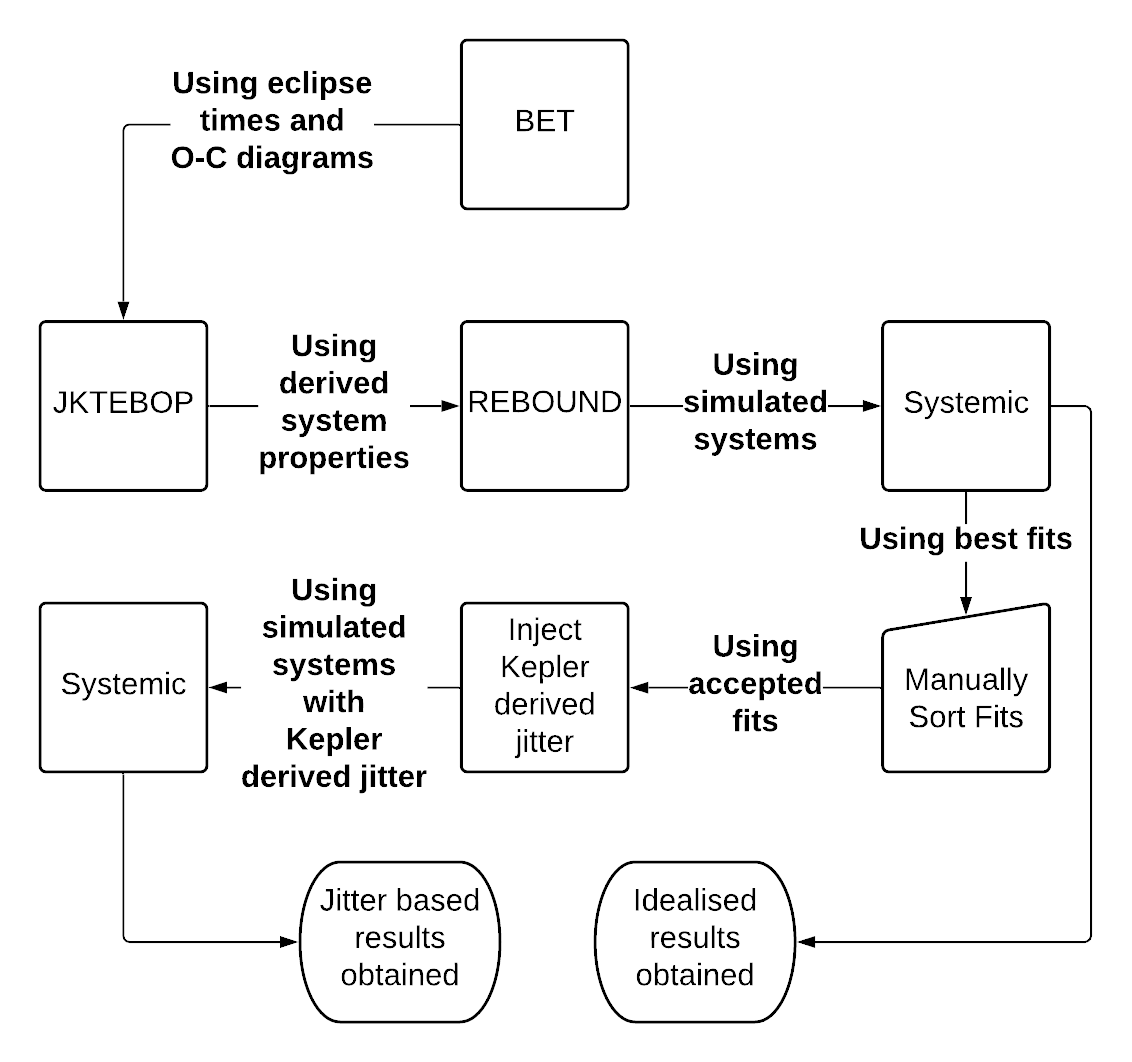}
    \caption{Summary of the major processes of the methodology. {\sc bet} was used to determine eclipse times and produce O-C diagrams to identify systems for use in the study. System properties were obtained from {\sc jktebop}, followed by simulated systems being made in {\sc rebound}. {\sc systemic} was then used to find the best fits (producing the idealised results). Kepler derived jitter was introduced into the simulated systems with visually comparable O-C fits and {\sc systemic} was again used to find the best fits (producing the jitter based results).}
    \label{fig:flowchart}
\end{figure}

The methodology has been summarised in the flow chart in Fig. \ref{fig:flowchart}.

\begin{figure}
    \includegraphics[width=\linewidth]{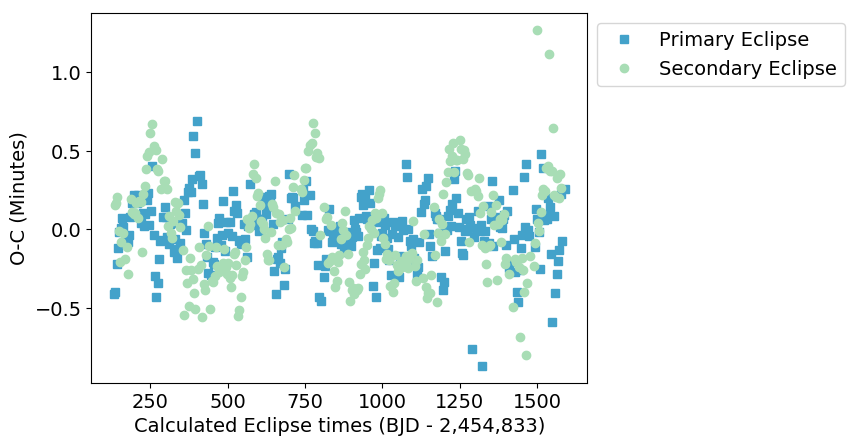}
    \caption{O-C diagram for KIC 6521542. KIC 6521542 was chosen as one of the base systems for this ETV study due to the quasi-periodic variations with typical amplitude between $\pm1$ minute.}
    \label{fig:6521542_oc}
\end{figure}

\begin{figure}
    \includegraphics[width=\linewidth]{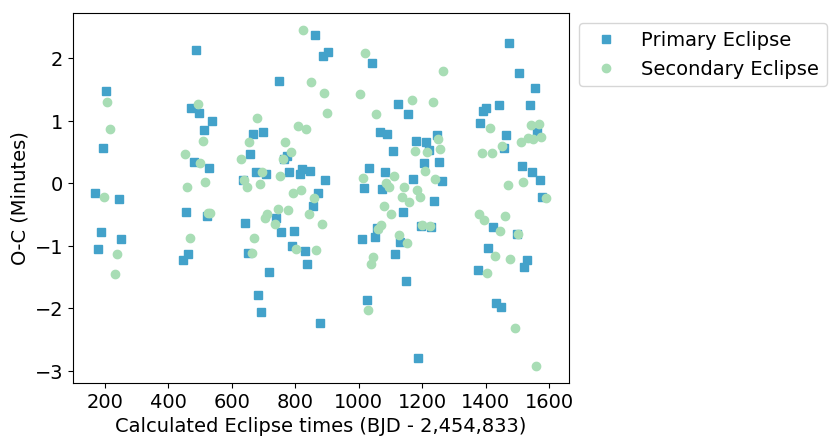}
    \caption{O-C diagram for KIC 3654950. KIC 3654950 was chosen as one of the base systems for this ETV study due to the random variations with typical amplitude between $\pm2.5$ minutes.}
    \label{fig:3654950_oc}
\end{figure}

\begin{figure}
    \includegraphics[width=\linewidth]{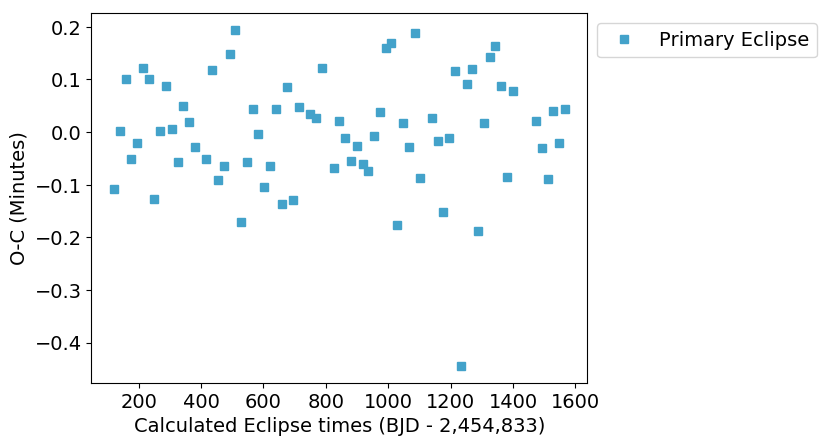}
    \caption{O-C diagram for KIC 6593363. KIC 6593363 was chosen as one of the base systems for this ETV study due to the random variations with typical amplitude between $\pm0.2$ minutes.}
    \label{fig:6593363_oc}
\end{figure}

The eclipse times of the eclipsing binary stars found in the 'Kepler Eclipsing Binary Star Catalog' were determined using a custom program, {\sc bet}, based on the software {\sc transit analysis package} or {\sc tap} \citep{gazak2012}. Three case study example systems, KIC 3654950, KIC 6521542 and KIC 6593363, were chosen as the basis of this study as the eclipse time variations appear minimal and random or quasi-periodic at most. The O-C diagrams for these systems can be seen in Figs. \ref{fig:6521542_oc}, \ref{fig:3654950_oc} and \ref{fig:6593363_oc}. It can be seen that variations range from a fraction of a minute (KIC 6593363) up to two minutes (KIC 3654950). If the variations were periodic it's possible that a third body would already be present in the system \citep{beuermann2010} and interfere with the results of the third body detection methods. The eclipsing binaries found in the Kepler eclipsing binary catalog have O-C diagram's with varying characteristics. KIC 3654950, KIC 6521542 and KIC 6593363 were selected as their O-C diagrams, with minimal, random and/or quasi-periodic variations, are also representative of the other O-C diagram characteristics seen from Kepler eclipsing binaries.

{\sc jktebop} \citep{southworth2004} was used to determine estimates for the characteristics of the binary star (including the mass ratio of the binary stars, orbital period, inclination). The temperature of the systems were estimated as in \cite{getley2017} and \cite{getley2020}. The temperatures were then used to estimate individual star masses in the binary system. These systems were also chosen for their different binary star orbital periods. The orbital periods for KIC 6521542, 3654950 and 6593363 are 4.42575, 8.13475 and 18.52783 days respectively. These complementary systems are thus used to determine how differing binary configurations alter the detectability of a third body.

{\sc rebound} is an N-body integrator with Python and C implementations \citep{rein2015}. Systems of bodies are able to be set up and integrated over time to determine eclipse timing variations for the characteristics of the objects entered. With the characteristics of the Kepler systems determined from {\sc jktebop} (Table ~\ref{tab:binarysystem}), these values were used to set up base binary star systems in {\sc rebound}. A series of third bodies was then added to each of the systems. The characteristics of the injected third body had masses ranging between 0.5 Jupiter masses and 500 Jupiter masses\footnote{Masses used were: 0.5, 1, 2, 3, 4, 5, 6, 7, 8, 9, 10, 15, 20, 25, 30, 35, 40, 45, 50, 55, 60, 65, 70, 75, 80, 85, 90, 95, 100, 110, 120, 130, 140, 150, 200, 300, 400 and 500 Jupiter masses}. The orbital period of the third body was set between 6 and 2000 days\footnote{Periods used were: 6, 7, 8, 9, 10, 15, 20, 25, 30, 35, 40, 45, 50, 100, 200, 300, 400, 500, 1000 and 2000 days}. Eccentricities were also set to 0.0, 0.1 and 0.5 for each mass/period combination along with random mean anomolies, longitude of ascending node and longitude of pericentre. The inclination of the third bodies was fixed to 70 degrees. If the inclination is any closer to 90 degrees, the third bodies will start to transit the parent stars and will be detectable via other methods \citep{charbonneau2006}. Therefore, an inclination of 70 degrees is more representative of an eclipse timing study system. These systems were then integrated with {\sc rebound} and the eclipse times of the binary stars were recorded. These simulated eclipse times form the basis of an idealised scenario i.e. no jitter was introduced due to unwanted internal effects, such as star spots, or external effects such as observational errors. O-C variations, or Kepler derived jitter, from the case study example systems can then be added to the simulated eclipse times. These eclipse times with Kepler derived jitter added then form the basis for a "real-world" scenario based on actual observation data.

\begin{table}
    \centering
    \caption{Properties for the three case study example systems used as the base of this eclipse timing study. Binary stars with varying orbital periods were selected in order to determine the effect binary orbital period has on the third body detection rate.}
    \label{tab:binarysystem}
    \begin{tabular}{lccc}
    \hline
    Property & KIC 6521542 & KIC 3654950 & KIC 6593363\\
    \hline
    Kepler $T_{eff}$ (K) & 5880 & 5233 & 5865 \\
    Kepler Mag & 14.280 & 15.858 & 12.893 \\
    Primary Mass ($M_{\odot}$) & 1.07 & 0.86 & 1.07 \\
    Secondary Mass ($M_{\odot}$) & 0.365 & 0.442 & 0.740 \\
    Period (d) & 4.42575 & 8.13475 & 18.52783 \\
    Inclination (\textdegree) & 88.99 & 89.25 & 89.93 \\
    \hline
    \end{tabular}
\end{table}

The binary star systems were set up in {\sc systemic} \citep{meschiari2009, meschiari2010} using the characteristics of the system found from {\sc jktebop} and the simulated eclipse times from {\sc rebound}. We followed an iterative process to determine the best possible fit. A third body was then inserted in to {\sc systemic} with the mass set in the range 5, 10, 25, 50, 100, 250, 500 and 1000 Jupiter Masses as initial values. The system properties were then optimised to search for the best fit values for the third body characteristics. The best two masses were selected as upper and lower limits and the systems were re-run to find the best fit. This was repeated until the minimum and maximum masses differ by less than a Jupiter mass. The best fit at the end of the calculations were then saved. O-C diagrams for the best fits from {\sc systemic} were then saved and sorted in to three categories: good fit, bad fit, uncertain fit. Manual intervention in finding a model with {\sc systemic} may be able to provide better fits, however given the extremely large number of systems that there are to work with and that one of the purposes of this investigation is to find the limits of a largely automated calculation, manual intervention is not practical.

After the systems were processed in {\sc systemic}, the next step was to determine if the fitted third body characteristics matched the injected third body characteristics. The O-C diagram of the simulated system vs the best-fit was inspected, if the O-C diagrams were a visually poor fit the system was rejected as not a detection. From here, the third body masses were checked to see if they fell within $\pm 25\%$, $\pm 50\%$ or $\pm 100\%$ of the injected third body's characteristics. Successful detections can then be used to determine what third body characteristics are detectable under ideal conditions as well as "real-world" conditions.

Given the large number of simulated systems, the long processing time and limited computing resources available some compromises had to be made. As the period of a third body can be estimated from the period of variability in the O-C diagram, the period of the third body in {\sc systemic} was fixed to the known simulated/injected period of the third body. This significantly reduced computing time required for fitting. After the entire set of idealised simulations were run, only systems that had visually acceptable O-C fits were then used for the simulated systems with Kepler derived jitter added. This is because
adding noise makes a detection less likely, therefore if a detection is unsuccessful under idealised conditions, it will be unsuccessful under less than idealised conditions.

\section{Results}

While finding the precise mass of an object is the ideal outcome, uncertainty is unavoidable. As such, we analyse the results with varying uncertainty to describe a mass detection. We considered the cases where the found mass was within $\pm 25\%$, $\pm 50\%$, $\pm 100\%$ of the injected third body's known mass. We also consider the effect of eccentricity and host binary star characteristics on detection rates.

Simulations that had best fit O-C variations that didn't visually match the actual O-C variations were immediately regarded as a non-detection. For the purposes of this study any O-C fit that was considered visually uncertain (i.e. it wasn't an obvious rejection) were included with the visually good O-C diagram fits in consideration as a possible detection. By including the systems with O-C diagrams that were deemed uncertain, we aim to remove some of the "human error" involved in sorting the O-C diagrams and letting the rest of the processes determine what was and was not a detection. 

\begin{figure*}
    \includegraphics[width=\linewidth]{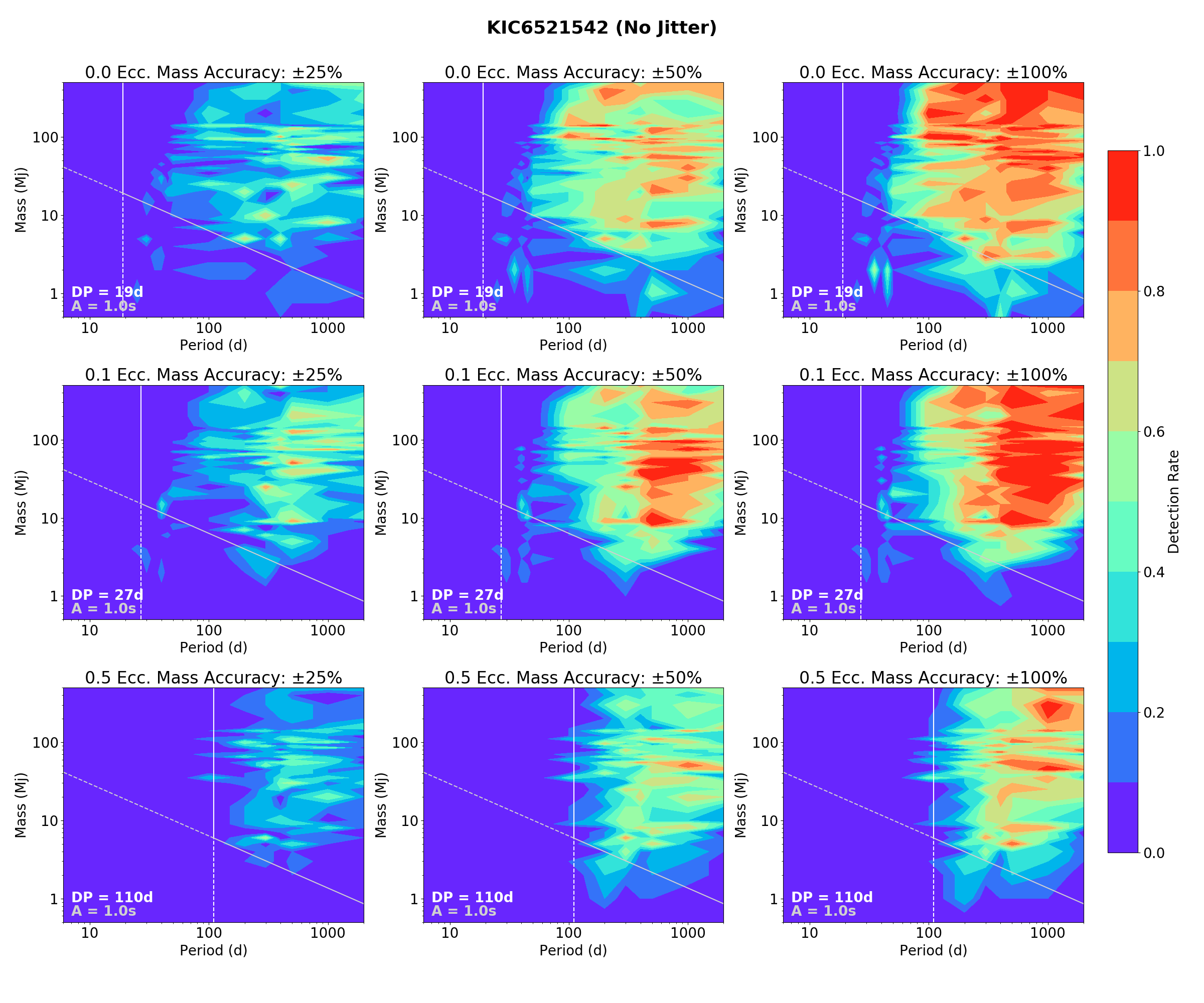}
    \caption{Detection rate of injected third bodies around KIC 6521542. Detections made under idealised conditions (i.e. with no Kepler derived jitter added). From top to bottom are third bodies injected with eccentricities (Ecc) of zero, 0.1 and 0.5 respectively. From left to right, mass accuracy used for a successful detection is $\pm 25\% $, $\pm 50\%$ and $\pm 100\%$. The minimum detection period (from eq. \protect{\ref{eq:detection_period}}) is shown by the vertical line. The mass-period relation ship (from eq. \protect{\ref{eq:sybilski}}) is shown by the diagonal line. The timing accuracy, A, was set to 1.0s. The solid portion of the lines indicate masses and/or periods that fit both equations \protect{\ref{eq:detection_period}} and \protect{\ref{eq:sybilski}}. The dashed portion indicate masses and/or periods that fit only one of the equations.}
    \label{fig:kic6521542-ideal}
\end{figure*}

\begin{figure*}
    \includegraphics[width=\linewidth]{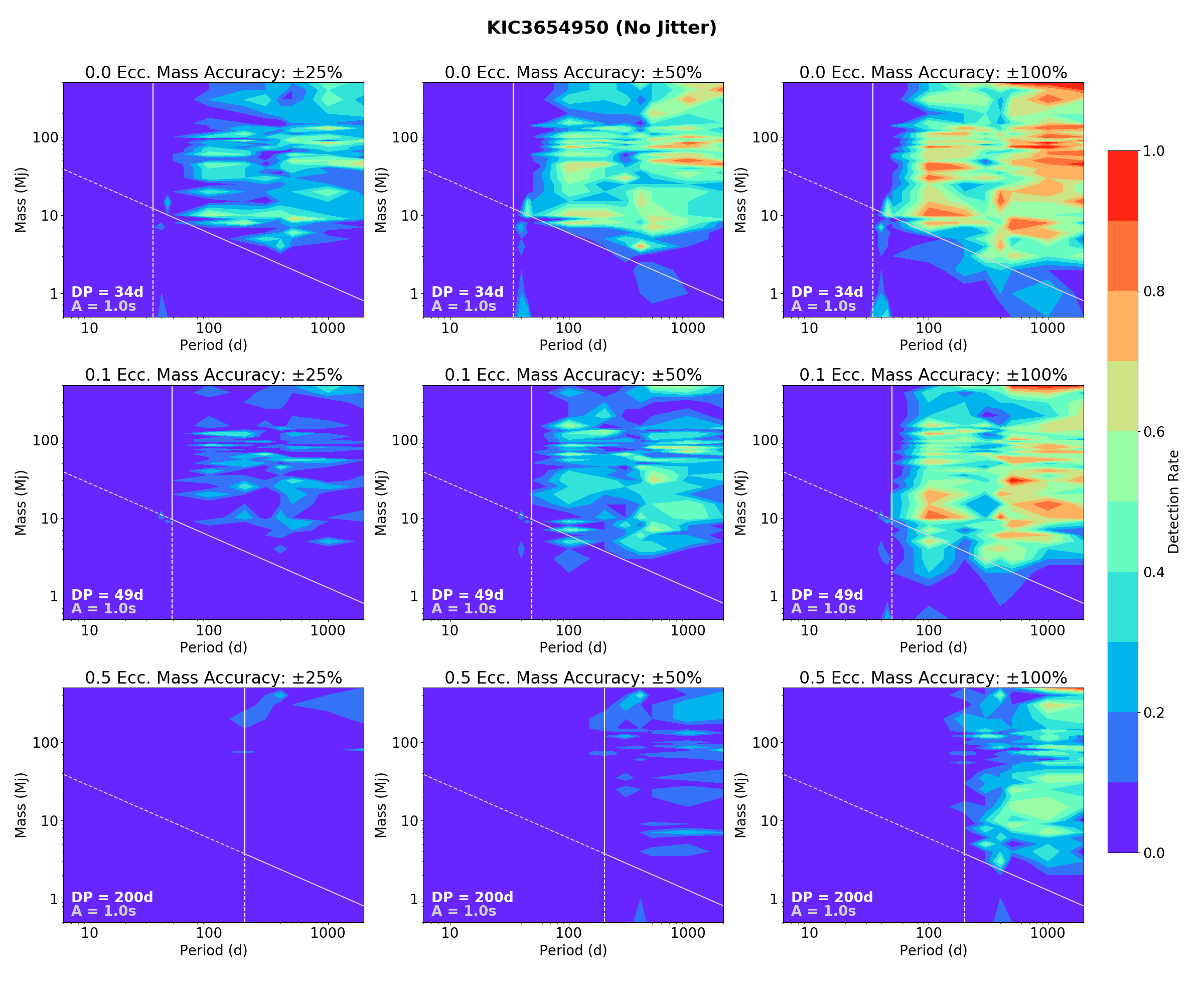}
    \caption{Detection rate of injected third bodies around KIC 3654950. Detections made under idealised conditions (i.e. with no Kepler derived jitter added). From top to bottom are third bodies injected with eccentricities (Ecc) of zero, 0.1 and 0.5 respectively. From left to right, mass accuracy used for a successful detection is $\pm 25\% $, $\pm 50\%$ and $\pm 100\%$. The minimum detection period (from eq. \protect{\ref{eq:detection_period}}) is shown by the vertical line. The mass-period relation ship (from eq. \protect{\ref{eq:sybilski}}) is shown by the diagonal line. The timing accuracy, A, was set to 1.0s. The solid portion of the lines indicate masses and/or periods that fit both equations \protect{\ref{eq:detection_period}} and \protect{\ref{eq:sybilski}}. The dashed portion indicate masses and/or periods that fit only one of the equations.}
    \label{fig:kic3654950-ideal}
\end{figure*}

\begin{figure*}
    \includegraphics[width=\linewidth]{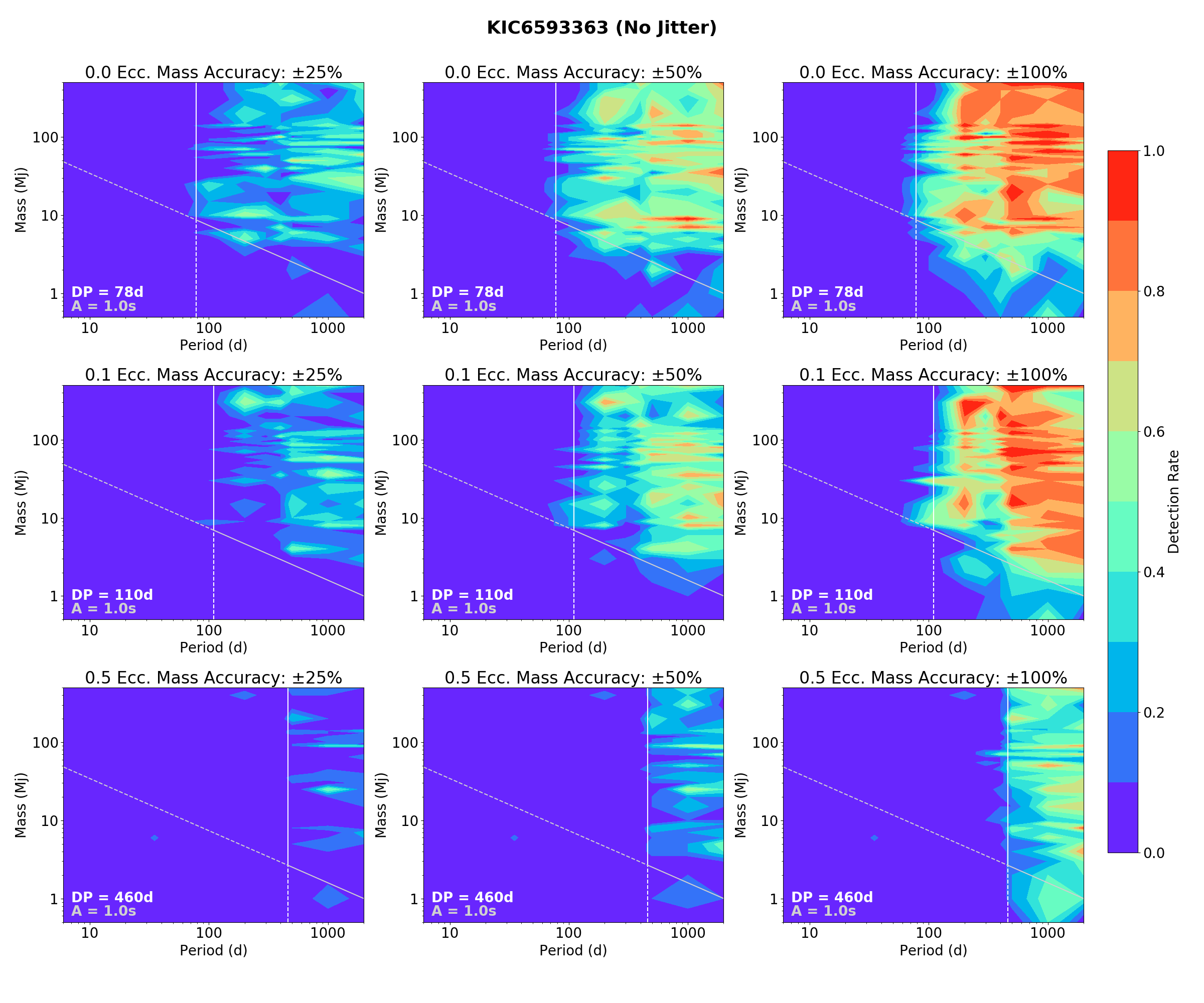}
    \caption{Detection rate of injected third bodies around KIC 6593363. Detections made under idealised conditions (i.e. with no Kepler derived jitter added). From top to bottom are third bodies injected with eccentricities (Ecc) of zero, 0.1 and 0.5 respectively. From left to right, mass accuracy used for a successful detection is $\pm 25\% $, $\pm 50\%$ and $\pm 100\%$. The minimum detection period (from eq. \protect{\ref{eq:detection_period}}) is shown by the vertical line. The mass-period relation ship (from eq. \protect{\ref{eq:sybilski}}) is shown by the diagonal line. The timing accuracy, A, was set to 1.0s. The solid portion of the lines indicate masses and/or periods that fit both equations \protect{\ref{eq:detection_period}} and \protect{\ref{eq:sybilski}}. The dashed portion indicate masses and/or periods that fit only one of the equations.}
    \label{fig:kic6593363-ideal}
\end{figure*}

\begin{figure*}
    \includegraphics[width=\linewidth]{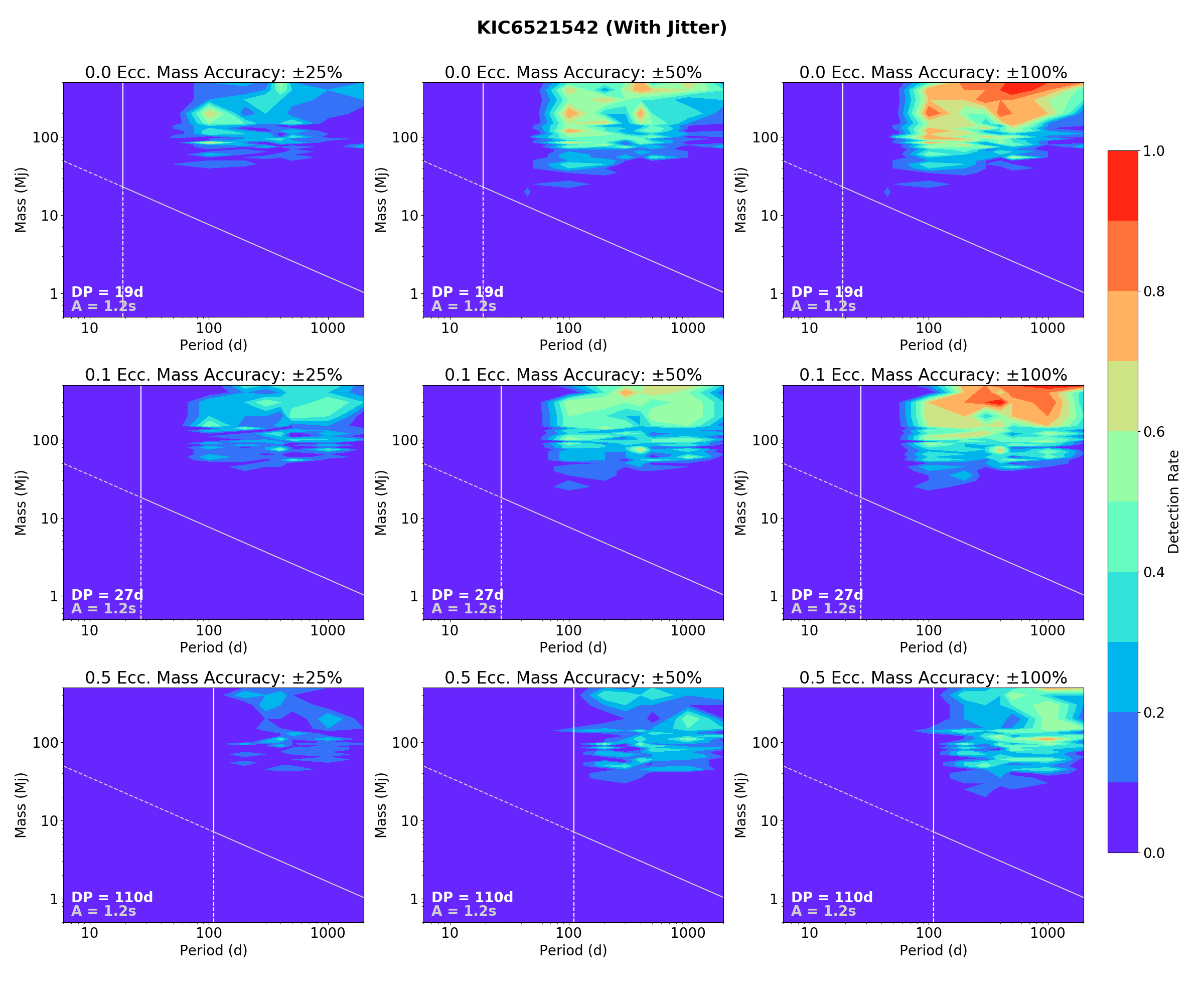}
    \caption{Detection rate of injected third bodies around KIC 6521542. Detections made under less than idealised conditions (i.e. with Kepler derived jitter added). From top to bottom are third bodies injected with eccentricities (Ecc) of zero, 0.1 and 0.5 respectively. From left to right, mass accuracy used for a successful detection is $\pm 25\% $, $\pm 50\%$ and $\pm 100\%$. The minimum detection period (from eq. \protect{\ref{eq:detection_period}}) is shown by the vertical line. The mass-period relation ship (from eq. \protect{\ref{eq:sybilski}}) is shown by the diagonal line. The timing accuracy, A, was estimated to be 1.2s. The solid portion of the lines indicate masses and/or periods that fit both equations \protect{\ref{eq:detection_period}} and \protect{\ref{eq:sybilski}}. The dashed portion indicate masses and/or periods that fit only one of the equations.}
    \label{fig:kic6521542-variations}
\end{figure*}

\begin{figure*}
    \includegraphics[width=\linewidth]{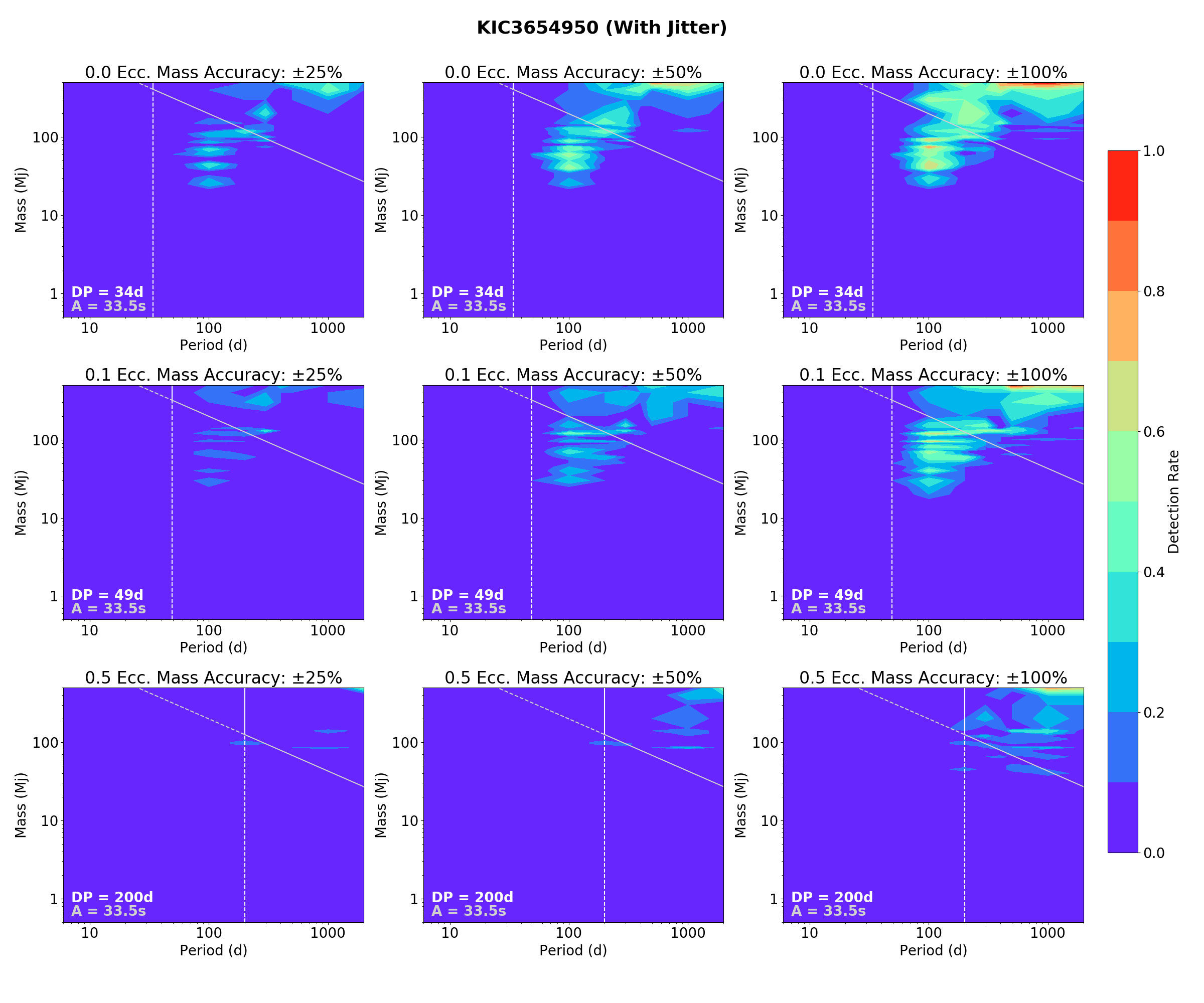}
    \caption{Detection rate of injected third bodies around KIC 3654950. Detections made under less than idealised conditions (i.e. with Kepler derived jitter added). From top to bottom are third bodies injected with eccentricities (Ecc) of zero, 0.1 and 0.5 respectively. From left to right, mass accuracy used for a successful detection is $\pm 25\% $, $\pm 50\%$ and $\pm 100\%$. The minimum detection period (from eq. \protect{\ref{eq:detection_period}}) is shown by the vertical line. The mass-period relation ship (from eq. \protect{\ref{eq:sybilski}}) is shown by the diagonal line. The timing accuracy, A, was estimated to be 33.5s. The solid portion of the lines indicate masses and/or periods that fit both equations \protect{\ref{eq:detection_period}} and \protect{\ref{eq:sybilski}}. The dashed portion indicate masses and/or periods that fit only one of the equations.}
    \label{fig:kic3654950-variations}
\end{figure*}

\begin{figure*}
    \includegraphics[width=\linewidth]{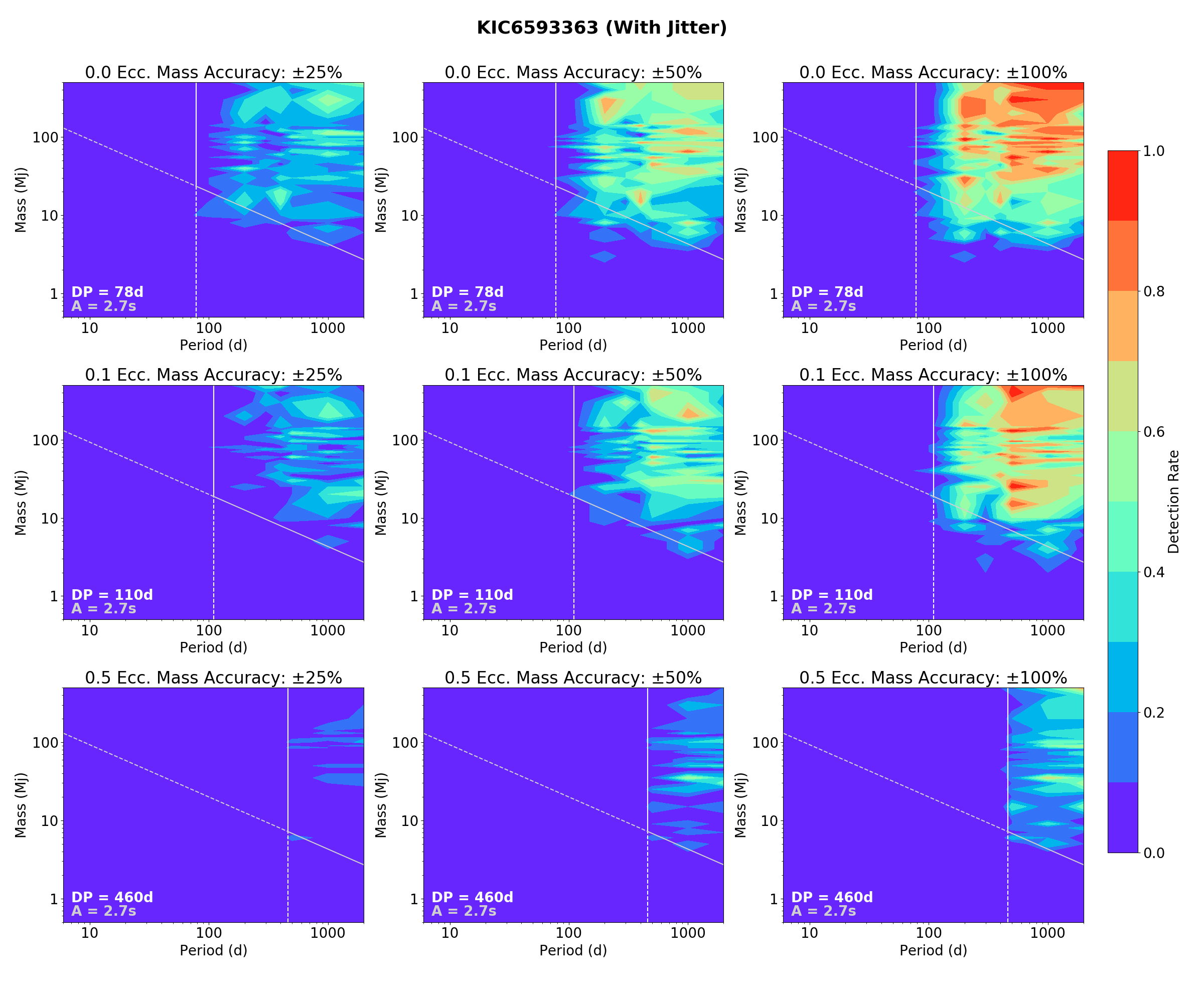}
    \caption{Detection rate of injected third bodies around KIC 6593363. Detections made under less than idealised conditions (i.e. with Kepler derived jitter added). From top to bottom are third bodies injected with eccentricities (Ecc) of zero, 0.1 and 0.5 respectively. From left to right, mass accuracy used for a successful detection is $\pm 25\% $, $\pm 50\%$ and $\pm 100\%$. The minimum detection period (from eq. \protect{\ref{eq:detection_period}}) is shown by the vertical line. The mass-period relation ship (from eq. \protect{\ref{eq:sybilski}}) is shown by the diagonal line. The timing accuracy, A, was estimated to be 2.7s. The solid portion of the lines indicate masses and/or periods that fit both equations \protect{\ref{eq:detection_period}} and \protect{\ref{eq:sybilski}}. The dashed portion indicate masses and/or periods that fit only one of the equations.}
    \label{fig:kic6593363-variations}
\end{figure*}

It is unsurprising that increasing the range of acceptable masses considered to be a detection results in increases in the detection rate (seen by comparing neighbouring columns in Figures ~\ref{fig:kic6521542-ideal} to \ref{fig:kic6593363-variations}). Adding Kepler derived jitter lowered the detection rate (for example by comparing Figures ~\ref{fig:kic3654950-ideal} and ~\ref{fig:kic3654950-variations}). However, it was also found that in all cases (with and without the introduced jitter), increasing the eccentricity of the third body lowered the successful detection rate of the third body.

The properties of the host binary star can have a noticeable effect on the detection rate of third bodies when the period of the third body is closer to the period of the host binary star. For example, comparing the detection rate of KIC6521542 (with a binary period of 4.42575 days) in Figures ~\ref{fig:kic6521542-ideal} and ~\ref{fig:kic6521542-variations} with the detection rate of KIC6593363 (that has a binary period of 18.52783 days) in Figures ~\ref{fig:kic6593363-ideal} and ~\ref{fig:kic6593363-variations}. With a longer host binary period, the detections occur at longer periods while a shorter host binary period has detections at shorter periods.

Smaller period changes in the host binary period may not lead to an entire shift in the period of detections. Smaller period changes may result in new regions where the number of detections drop significantly or even completely (figures ~\ref{fig:kic6521542-ideal} and ~\ref{fig:kic3654950-ideal}).

By comparing the best-fit mass with the actual mass of the injected third body (Table \ref{tab:over_under_ideal} and Table \ref{tab:over_under_variations}), we find that generally we are more likely to over estimate the injected third body's mass than under estimate the mass. This is particularly evident in the $\pm 100\%$ accuracy for both the idealised results and simulations injected with Kepler derived jitter. When considering systems with Kepler derived jitter injected at $\pm 25\%$ there are a similar number of systems where the third body mass is under estimated as over estimated.

\begin{table}
    \centering
    \caption{The total number of successful detections and the number of systems that under estimated and over estimated the third body mass at each eccentricity (Ecc), accuracy level and system in the study for the simulations with no variations added.}
    \begin{tabular}{c|c|c|c|c|c}
        \hline
         System (KIC) & Ecc & Accuracy & \# under & \# over & Total \\
         \hline
         \multirow{9}{4em}{6521542} & \multirow{3}{*}{0.0} & $\pm 25\%$ & 330 & 503 & 833\\
         &  & $\pm 50\%$ & 646 & 964 & 1610\\
         &  & $\pm 100\%$ & 770 & 1261 & 2031\\
         \cline{2-6}
         & \multirow{3}{*}{0.1} & $\pm 25\%$ & 248 & 630 & 878 \\
         &  & $\pm 50\%$ & 374 & 1170 & 1544 \\
         &  & $\pm 100\%$ & 416 & 1436 & 1852 \\
         \cline{2-6}
         & \multirow{3}{*}{0.5} & $\pm 25\%$ & 104 & 368 & 472\\
         &  & $\pm 50\%$ & 168 & 794 & 962 \\
         &  & $\pm 100\%$ & 189 & 1034 & 1223 \\
         \hline
         \multirow{9}{4em}{3654950} & \multirow{3}{*}{0.0} & $\pm 25\%$ & 273 & 354 & 627\\
         &  & $\pm 50\%$ & 445 & 612 & 1057\\
         &  & $\pm 100\%$ & 470 & 1061 & 1531\\
         \cline{2-6}
         & \multirow{3}{*}{0.1} & $\pm 25\%$ & 117 & 187 & 304 \\
         &  & $\pm 50\%$ & 191 & 442 & 633 \\
         &  & $\pm 100\%$ & 203 & 1052 & 1255 \\
         \cline{2-6}
         & \multirow{3}{*}{0.5} & $\pm 25\%$ & 18 & 49 & 67\\
         &  & $\pm 50\%$ & 33 & 158 & 191 \\
         &  & $\pm 100\%$ & 35 & 490 & 525 \\
         \hline
         \multirow{9}{4em}{6593363} & \multirow{3}{*}{0.0} & $\pm 25\%$ & 276 & 302 & 578\\
         &  & $\pm 50\%$ & 532 & 624 & 1156\\
         &  & $\pm 100\%$ & 618 & 1080 & 1698\\
         \cline{2-6}
         & \multirow{3}{*}{0.1} & $\pm 25\%$ & 124 & 303 & 427 \\
         &  & $\pm 50\%$ & 192 & 686 & 878 \\
         &  & $\pm 100\%$ & 213 & 1335 & 1548 \\
         \cline{2-6}
         & \multirow{3}{*}{0.5} & $\pm 25\%$ & 44 & 74 & 118\\
         &  & $\pm 50\%$ & 57 & 200 & 257 \\
         &  & $\pm 100\%$ & 63 & 470 & 533 \\
         \hline
    \end{tabular}
    \label{tab:over_under_ideal}
\end{table}

\begin{table}
    \centering
    \caption{The total number of successful detections and the number of systems that under estimated and over estimated the third body mass at each eccentricity (Ecc), accuracy level and system in the study for the simulations with Kepler variations added.}
    \begin{tabular}{c|c|c|c|c|c}
        \hline
         System (KIC) & Ecc & Accuracy & \# under & \# over & Total \\
         \hline
         \multirow{9}{4em}{6521542} & \multirow{3}{*}{0.0} & $\pm 25\%$ & 126 & 134 & 260\\
         &  & $\pm 50\%$ & 253 & 223 & 476\\
         &  & $\pm 100\%$ & 296 & 353 & 649\\
         \cline{2-6}
         & \multirow{3}{*}{0.1} & $\pm 25\%$ & 109 & 156 & 265 \\
         &  & $\pm 50\%$ & 175 & 282 & 457 \\
         &  & $\pm 100\%$ & 195 & 429 & 624 \\
         \cline{2-6}
         & \multirow{3}{*}{0.5} & $\pm 25\%$ & 38 & 101 & 139\\
         &  & $\pm 50\%$ & 69 & 213 & 282 \\
         &  & $\pm 100\%$ & 84 & 332 & 416 \\
         \hline
         \multirow{9}{4em}{3654950} & \multirow{3}{*}{0.0} & $\pm 25\%$ & 62 & 59 & 121\\
         &  & $\pm 50\%$ & 108 & 113 & 221\\
         &  & $\pm 100\%$ & 114 & 235 & 349\\
         \cline{2-6}
         & \multirow{3}{*}{0.1} & $\pm 25\%$ & 36 & 45 & 81 \\
         &  & $\pm 50\%$ & 61 & 109 & 170 \\
         &  & $\pm 100\%$ & 64 & 252 & 316 \\
         \cline{2-6}
         & \multirow{3}{*}{0.5} & $\pm 25\%$ & 17 & 19 & 36\\
         &  & $\pm 50\%$ & 25 & 37 & 62 \\
         &  & $\pm 100\%$ & 25 & 110 & 135 \\
         \hline
         \multirow{9}{4em}{6593363} & \multirow{3}{*}{0.0} & $\pm 25\%$ & 211 & 247 & 458\\
         &  & $\pm 50\%$ & 380 & 493 & 873\\
         &  & $\pm 100\%$ & 436 & 827 & 1263\\
         \cline{2-6}
         & \multirow{3}{*}{0.1} & $\pm 25\%$ & 112 & 187 & 299 \\
         &  & $\pm 50\%$ & 177 & 436 & 613 \\
         &  & $\pm 100\%$ & 190 & 862 & 1052 \\
         \cline{2-6}
         & \multirow{3}{*}{0.5} & $\pm 25\%$ & 27 & 32 & 59\\
         &  & $\pm 50\%$ & 42 & 99 & 141 \\
         &  & $\pm 100\%$ & 48 & 208 & 256 \\
         \hline
    \end{tabular}
    \label{tab:over_under_variations}
\end{table}

\section{Discussion}

The truncation radius of a binary star is expected to range between 1.8 and 2.6 times the binary separation, $a_b$ \citep{pierens2007}. Using Kepler's third law ($P^2 \propto a^3$) this would be an approximate range of between 2.4 and 4.2 times the orbital period of the binary. For the three systems listed in Table ~\ref{tab:binarysystem} no third body detections would be expected with orbital periods less than approximately 34.1d, 18.6d and 77.8d for KIC 3654950, 6521542 and 6593363 respectively. As seen in Figures ~\ref{fig:kic6521542-ideal} to ~\ref{fig:kic6593363-variations} no detections were made within these ranges in either the idealised case (i.e. the Kepler mission time sampling with no added jitter) or the case where Kepler derived jitter was introduced. The smallest detection, in either case at any mass, was at 70d for KIC 3654950, 40d for KIC 6521542 and 100d for KIC 6593363. Detections increase significantly when a third body has an orbital period at or greater than 100d both in the ideal scenario and with variations. This indicates that while it is possible in some cases for Eclipse Timing Variation studies to detect third bodies close to the minimum formation period they are most sensitive to 100d+ orbital periods.

Using the approximate upper limit of the truncation radius as a foundation, and comparing the start of detections in both the idealised case and the case with Kepler derived jitter added for all eccentricities, we empirically find that the minimum detection period can be approximately described with the equation

\begin{equation}
    DP = (59{e_p}^2 + 12.1e_p + 4.2)P_b
    \label{eq:detection_period}
\end{equation}

where DP is the minimum detection period, $e_p$ is the eccentricity of third body and $P_b$ is the orbital period of the host binary. This detection period is indicated in Figs. \ref{fig:kic6521542-ideal} to \ref{fig:kic6593363-variations} by the solid and dashed vertical lines. As we simulated the third bodies with 20 discrete periods, the precise period where detections begin had to be estimated. However, the step sizes were 100 days or less for periods up to 500 days. As the detections all appear to begin at or before a 500 day orbital period, equation \ref{eq:detection_period} is expected to be a useful and accurate estimate. Should future studies use smaller step sizes, the equation may be able to be refined further.

From \cite{sybilski2010}, the equation to obtain a planet's mass from a period and timing amplitude is 

\begin{equation}
    M_P = \left(\frac{4{\pi}^2{M_B}^2}{{P}^2G} \right) ^{\frac{1}{3}}(Ac)
    \label{eq:sybilski}
\end{equation}

where $M_P$ and $P$ is the mass and period of a planet/third body companion, $M_B$ is the total mass of the binary star, $G$ is the gravitational constant, $A$ is the timing amplitude and $c$ is the speed of light. As such, with the total mass of the binary star and an estimate for the timing errors from a system, we can calculate the minimum detectable mass for a given orbital period of a third body companion. This is shown in Figs. \ref{fig:kic6521542-ideal} to \ref{fig:kic6593363-variations} by the solid and dashed vertical lines.

By using both equations \ref{eq:detection_period} and \ref{eq:sybilski} we can calculate an approximate minimum orbital period and an approximate minimum mass for detections of a third body companion at a specific orbital period. As such, reliable approximations on what type of companions may be detectable can be found from minimal binary star information. These realistic expectations can be used to as a guide for future eclipse timing studies.

From \cite{watson2010}, the amplitude of the timing deviation, $\delta t$, is related to the mass of the exterior planet, $M_p$, and its semi-major axis, $a_{out}$ by:

\begin{equation}
    \delta t \approx \left(\frac{M_p}{M_J}\right)\left(\frac{a_{out}}{au}\right)
    \label{eq:mass_period}
\end{equation}

As a result of equation \ref{eq:mass_period}, assuming the minimum detectable timing deviation, a detectable third body will have a decreasing mass as the orbital period increases. This can generally be seen to be the case, particularly in the ideal simulation scenario in Figures \ref{fig:kic6521542-ideal}, \ref{fig:kic3654950-ideal} and \ref{fig:kic6593363-ideal}. It can also be seen from Fig. \ref{fig:kic3654950-variations} that introducing jitter affects this property of timing deviations. As such, while simulations are a great launching point and can be used to rule out detections based on mass/period properties, this does not guarantee a detected third body is accurate.

The largest difference between the idealised systems and systems with Kepler derived jitter added can be seen between Fig. \ref{fig:kic3654950-ideal} and Fig. \ref{fig:kic3654950-variations} for KIC 3654950. There is a significant detection rate across a wide range of masses and periods when looking at idealised simulations alone. However, with the Kepler derived jitter added the detection rate is very low except for the larger brown dwarf and stellar masses at low eccentricity. Low detection rate occurs in the systems with shorter orbital periods, i.e. KIC 3654950 and 6521542. The estimated timing accuracy for KIC 3654950 at 33.5s is significantly larger than the other systems and explains the significant difference between the idealised systems and the simulations with Kepler derived jitter added. There are a number of detections below the timing accuracy line but only at shorter orbital periods. The reason for these detections is unclear.

In Fig. \ref{fig:kic3654950-ideal}, there appears to be an exception to the above. With 0.0 and 0.1 eccentricity, detections of 0.5 - 1$M_J$ third bodies occur between 35d and 50d orbital periods. This is particularly evident with $\pm 100\%$ accuracy but can also be seen in $\pm 25 \%$ and $\pm 50 \%$ accuracy with 0.0 eccentricity. The extra group of detections are not seen in other systems and not seen when Kepler derived jitter was included in the simulations. It is possible that these detections are just coincidences and not real detections. However, this raises the question of why the detections are grouped together and not randomly spread around the various simulated values.

It is also possible for the magnitude of the binary star to have an effect on detectability. The brighter the observed target, the better signal-to-noise ratio (SNR) that is obtainable and therefore less variability within the observations. With a magnitude of 15.858, KIC 3654950 is the dimmest of the three case study example systems. It is therefore likely that the poorer timing accuracy, and lack of detections, is partially the result of the increased magnitude.

\subsection{KIC 5095269}

KIC 5095269 has a 7.7$M_J$ planetary mass third body in a 237.70817d orbital period with an eccentricity of 0.06 \citep{getley2017}. The orbital period of the host binary is approximately 18.61d which closely matches orbital period of the binary found in KIC 6593363. From Fig. \ref{fig:kic6593363-variations} we can see the detection rate is very low (at 0.0 eccentricity) or zero (at 0.1 eccentricity) at this mass/period when a $\pm25\%$ mass accuracy is used but significantly higher (at 0.0 eccentricity) with a $\pm50\%$ mass accuracy or $\pm100\%$ (at 0.1 eccentricity). We can also use equations \ref{eq:detection_period} and \ref{eq:sybilski} to estimate the mass and periods in a detectable range. The minimum detectable period is estimated to be 96d. With an approximate median timing error of 2.4s, the approximate minimum detectable mass of a third body with a 237.70817d orbital period is 9.6$M_J$. From Fig. \ref{fig:kic6593363-variations} detections can be made with slightly lower masses than estimated from the timing error. As detection of a third body at this mass and period is possible, confidence of the third body's existence is increased. We can see from Table ~\ref{tab:over_under_variations} that the mass is more likely to be an over estimate than an under estimate. It is therefore possible the 7.7$M_J$ may be an upper estimate.

\subsection{KIC 7821010}

KIC 7821010 has a stable planetary third body with mass of $\sim 2.6M_J$ \citep{borkovits2016,getley2020}. The host binary has an orbital period of 24.238219d while the third body has an orbital period of 991d and an eccentricity of 0.372. The most comparable system with these properties is KIC 6593363 (Fig. \ref{fig:kic6593363-variations}). At 0.1 eccentricity a third body with these properties have a low detection rate, while at 0.5 eccentricity there are no detections. We have seen that detection rates increase as the host binary orbital period increases. Using equation \ref{eq:detection_period} we find that third bodies around KIC 7821010 have a minimum detection period of 400d. An approximate minimum detectable mass of a third body with an orbital period of 991d and an approximate median timing error of 0.18s is just 0.34$M_J$. Not only is this planetary mass third body expected to be detectable with an eclipse timing study but Saturn mass third bodies could be detectable using eclipse timing studies around KIC 7821010 (or similar systems) with more observations.

\subsection{Brown Dwarf Desert}

Brown dwarfs are generally considered to be bodies with masses ranging from approximately 13$M_J$ to 80$M_J$ \citep{spiegel2011,sahlmann2011}. There exists a brown dwarf desert where brown dwarf bodies are sporadically found around single stars or multiple star systems for a range of orbital periods \citep{grether2006,fontanive2019}. An estimated 16\% of Sun-like stars have third bodies in close orbits of less than 5 years. However, less than 1\% of these are brown dwarf masses.

Third bodies within the brown dwarf mass range with orbits less than $\sim$5 years are detectable around all systems listed in Table \ref{tab:binarysystem}. This supports the idea that the brown dwarf desert is not due to detection issues. Therefore, the lack of brown dwarfs are more likely due to other factors such as formation or migration processes as stated in \cite{grether2006}.

\subsection{Computing Resources}\label{sec:computing}

Ideally, more systems, more eccentricities and more datasets would have been included in this study. While it is clear the detection rate drops as the eccentricity of the third body increases, including more eccentricities would have allowed a clearer understanding of the detection rate. For example, is the drop relatively linear or is there a "detection rate cliff" where eccentricity has minimal effect and then rapidly has a significant effect? More systems would have allowed for a clearer understanding of the effect of the host binary orbital period on the detection rate for orbital periods between the short, 8d, systems and the longer, 18d, systems. The simulations are resource intensive to run and due to the technical limitations we chose to thoroughly cover a small number of representative systems rather than partially cover a larger number of systems. For example, 38 different masses, 20 different orbital periods, three different eccentricities each combination run with 10 random initial conditions for each of the three systems is a total of 68,400 simulations just for the ideal conditions. Each simulation required one CPU and had a wall-time of 24 hours. An additional system or eccentricity would therefore add a minimum of 22,800 simulations each and significantly increases the amount of resources needed. 

\subsection{Application to Other Space-Based Photometric Datasets}

The simulations results here, including equation \ref{eq:detection_period}, are specific to the Kepler dataset. Nevertheless, the approach taken here can be applied to other space-based photometric datasets of similar precision, cadence and extended time coverage. Thus we briefly consider the application of our approach to TESS and PLATO.

TESS observes sectors of the sky for 27.4 days, before re-pointing the field of view, with the all-sky survey taking 2 years \citep{ricker2014}. While TESS provides high precision observations of bright nearby stars, the duration of the observations of the objects is significantly shorter than Kepler's. Therefore, detecting third bodies with orbital periods greater than the 27.4d seems unlikely. So equations \ref{eq:detection_period} and \ref{eq:sybilski} can apply to missions such as TESS, although the maximum detectable period is then limited to the duration of the observations. In addition, as mentioned in section \ref{sec:computing}, the computational resources required to perform the necessary simulations are substantial. Consequently, we leave the exploration of the TESS dataset to the future when there is a more extended time coverage.

We also note the upcoming PLATO (PLAnetary Transits and Oscillations of stars) mission aims to observe bright stars (2 to 3 magnitudes brighter than Kepler observed stars) for a period of 4 years. By observing brighter stars than Kepler, PLATO measurements should have a greater precision than Kepler over similar time periods. As such it is feasible that our approach can be applied to the PLATO datasets.

\section{Summary and Conclusions}

In this paper, we have simulated the Kepler eclipsing binary systems KIC 3654950, KIC 6521542 and KIC 6593363 with injected third bodies with varying characteristics in order to determine 1) the detectability of third bodies with specific masses and periods, 2) the effect of "real-world" observations (or Kepler derived jitter with the Kepler mission time sampling) on the detectability of third bodies and 3) the effectiveness of using eclipse timing variations to hunt for planetary mass third bodies.

Our study finds that, when using empirical data from the Kepler Eclipsing Binary Star Catalog as a starting point, in an idealised situation (i.e. with no Kepler derived jitter), small mass third bodies are able to be detected at long orbital periods while large mass third bodies are able to be detected in shorter orbital periods. This agrees with the results of the simulation analysis performed by \cite{sybilski2010}. We however find that in less than idealised situations, i.e. with the Kepler derived jitter added to the simulations, that this property only holds true with larger binary orbital periods. We also find that the eccentricity of a third body has a significant effect on the detection rate of third bodies, with a larger eccentricity making detection significantly more difficult in less than idealised circumstances. The brightness of the observed target is also likely to play a role in the detectability of third bodies.

The truncation radius of a binary star is expected to be between approximately 2.4 and 4.2 times the orbital period of the binary \citep{pierens2007}. As such, no third bodies are expected to form within this radius and would only exist with planetary migration. We find that for binary stars with short orbital periods ($\sim 4d$ and $\sim 8d$) that third bodies aren't detected until approximately twice the truncation radius. However, when the binary star has a longer orbital period such as KIC 6593363 with an 18.52783d orbital period third bodies can be detected close to the truncation radius. With this information we are able to see the detection rate of a Kepler observed system can be bound by equations \ref{eq:detection_period} and \ref{eq:sybilski}.

It can be seen from KIC 6593363 that detection rates significantly increase with the longer host binary orbital period. We draw the conclusion that with the longer orbital period, and thus greater separation between the stellar components of the binary that the stars are not only significantly detached but also at a great enough distance that tidal distortions are less prominent. This results in smaller variations from within the system itself and allows the detection rate to increase.

Eclipse time variation studies can be, and have been, used to find planetary mass third bodies \citep{borkovits2016,getley2020}. However, ETV studies are more sensitive to brown dwarf and stellar mass companions. As such, while ETV studies are an important tool in finding third bodies (of a wide range of masses and periods) it's important to understand the potential limitations of such studies in order to guide expectations and maximise the use of resources. As more missions like Kepler are launched, existing binary systems will be observed for even longer periods of time. This will allow third bodies with longer orbital periods and smaller masses to be discovered through ETV studies.

The upcoming PLATO mission aims to obtain high precision observations of bright stars for similar time periods to Kepler \citep{catala2009}. By observing bright stars, photon noise sources can be kept to a minimum. As such, PLATO observed binary stars are good potential targets for an ETV study.

\section*{Acknowledgements}

This research has been supported by an Australian Government Research Training Program Scholarship. This research was undertaken with the assistance of resources from the National Computational Infrastructure (NCI Australia), an NCRIS enabled capability supported by the Australian Government.

We would like to thank the referee and editor for their helpful comments.

\section*{Data availability}

The data underlying this article will be shared on reasonable request to the corresponding author.




\bibliographystyle{mnras}
\bibliography{main} 


\bsp	
\label{lastpage}
\end{document}